\documentclass[a4paper,11pt]{article}
\usepackage{jcappub} 
\usepackage{lineno}

\title{\boldmath Bose–Einstein condensation and dark radiation production from bosonic warm dark matter}

\author{Zhijian Zhang}
\affiliation{School of Physics and Astronomy,
Beijing Normal University,
Beijing 100875, China}

\emailAdd{zhangzj221@mail.bnu.edu.cn}

\abstract{We investigate the quantum statistical effects of bosonic warm dark matter (BWDM) in a microscopic model in which dark radiation (DR) is produced through the annihilation process $\chi\chi\rightarrow\phi\phi$ mediated by a Yukawa interaction. By decomposing the BWDM phase-space distribution into thermal and Bose--Einstein condensed components, we derive the thermally averaged annihilation cross sections for the thermal--thermal, thermal--condensate, and condensate--condensate channels, including the Sommerfeld enhancement. We find that, for the relatively small scalar DR masses considered here, the thermal--condensate channel remains subdominant. The transition between the thermal--thermal dominated and condensate--condensate dominated regimes is characterized by a critical condensate fraction $r_{\rm c}$, which depends mainly on the BWDM mass and the scalar DR mass. For condensate fractions above this critical value, the condensate--condensate channel can dominate the present-day annihilation rate. We further constrain the BWDM parameter space using the observational upper bound on the additional effective number of relativistic species, $\Delta N_{\mathrm{eff}}$.}

\begin{document}
\maketitle
\flushbottom
\section{Introduction}

Currently, several open questions remain in the minimal cosmological constant plus cold dark matter ($\Lambda$CDM) scenario. One of the most prominent is the Hubble tension, namely the discrepancy between the Hubble constant inferred from local distance measurements, $H_0=73.04\pm1.04\,\mathrm{km\,s^{-1}\,Mpc^{-1}}$, and that inferred from cosmic microwave background (CMB) observations within the standard $\Lambda$CDM model, $H_0=67.4\pm0.5\,\mathrm{km\,s^{-1}\,Mpc^{-1}}$ \cite{Planck:2018,Verde:2019,DiValentino:2021,Schoneberg:2022,Riess:2022}. This discrepancy has motivated extensive investigations of extensions to the standard cosmological model. Meanwhile, the nature and microscopic properties of dark matter (DM), which constitutes a substantial fraction of the matter content of the Universe, remain unknown. Various DM candidates and production mechanisms have therefore been extensively investigated \cite{Dodelson:1994,Peebles:2000,Bode:2001,Barkana:2001,Dolgov:2002,Sikivie:2010,Angulo:2013,Hui:2016,Aboubrahim:2020,Liang:2020,Sommer:2024,Mondal:2025,Baryakhtar:2025}.

One possibility explored in attempts to alleviate the Hubble tension is the existence of additional relativistic degrees of freedom, commonly referred to as dark radiation (DR) \cite{CyrRacine:2013,BuenAbad:2015,Tang:2016,Choi:2020,Saravanan:2025,Bagherian:2025,BuenAbad:2025,Garny:2026,Chang:2026}. Their energy density is conventionally characterized by the effective number of relativistic species, $N_{\mathrm{eff}}$, or its deviation $\Delta N_{\mathrm{eff}}$ from the standard cosmological prediction. Additional radiation can modify the expansion rate before recombination and affect the inferred value of $H_0$. However, its ability to alleviate the Hubble tension depends on the properties and interactions of the DR component, and additional radiation alone does not necessarily resolve the discrepancy \cite{Archidiacono:2013,DiValentino:2021,Gariazzo:2023,BuenAbad:2024,Allali:2025,Mescia:2026,Garny:2026}. Moreover, the contribution of additional relativistic species to the expansion history and the evolution of cosmological perturbations leaves observable imprints on the CMB, providing a means of constraining their abundance. These considerations motivate investigations of the microscopic origin of DR and the mechanisms through which it can be produced in the dark sector.

The dark sectors may contain non-gravitational interactions between DM and other dark-sector particles. Such interactions have been investigated in various cosmological scenarios, including models involving DM--DR scattering and interacting dark radiation \cite{BuenAbad:2015,Tang:2016,Choi:2020,Bagherian:2025,BuenAbad:2025,Garny:2026}. These interactions can modify the evolution of DM and DR and leave observable signatures in cosmological data \cite{Brinckmann:2023,Bagherian:2025,BuenAbad:2025}. In particular, DM annihilation into relativistic particles provides a microscopic mechanism for transferring energy from DM to DR. The resulting DR abundance depends on the annihilation rate and the cosmological evolution of the interacting components. In this work, we investigate DR production through DM annihilation during the epoch after matter--radiation equality and constrain the resulting abundance using observational bounds on $\Delta N_{\mathrm{eff}}$.

Bosonic warm dark matter (BWDM) provides a particularly interesting framework in which quantum statistical effects can influence microscopic interactions. For sufficiently cold and highly occupied bosonic particles, Bose--Einstein condensation (BEC) can lead to a macroscopic occupation of the zero-momentum ground state \cite{Sin:1994,Sikivie:2009,Hui:2017,Mondal:2024,Mondal:2025,HaghanI:2026}. In a mixed thermal--condensate state, the thermal and condensate components have distinct phase-space distributions and can therefore contribute differently to particle annihilation. The condensate fraction consequently provides an additional parameter controlling the relative importance of the different annihilation channels. Understanding how this quantum-statistical structure affects the production of DR is the main motivation for the present study.

Furthermore, when DM particles interact through a light scalar mediator, the resulting attractive Yukawa potential can give rise to Sommerfeld enhancement, modifying the annihilation rate at low velocities. For a mediator with finite mass, the enhancement saturates in the sufficiently low-velocity regime. Since particles in a BEC occupy the zero-momentum state, condensate--condensate annihilation probes this saturated regime, whereas the thermal channels involve continuous momentum distributions. The BEC fraction determines the relative number densities of the thermal and condensate components, while their different phase-space distributions lead to different thermally averaged annihilation rates across the three channels, including the effects of Sommerfeld enhancement. Their combined effect determines the dominant annihilation channel and influences the energy transfer from BWDM to DR.

Motivated by these considerations, we investigate the quantum statistical effects of BWDM in a microscopic model where DR is produced through the annihilation process $\chi\chi\rightarrow\phi\phi$. Here, $\chi$ denotes the BWDM particle, while $\phi$ is a light scalar field whose relativistic produced particles constitute the DR component. The interaction between the BWDM particles is described by a Yukawa potential, and the Sommerfeld enhancement is included in the annihilation processes.

By decomposing the BWDM phase-space distribution into thermal and condensate components, we calculate the thermally averaged annihilation cross sections for the thermal--thermal, thermal--condensate, and condensate--condensate channels. We investigate how the condensate fraction changes the relative contributions of these channels and determines the transition between different annihilation-dominated regimes, characterized by a critical condensate fraction $r_{\rm c}$. We then calculate the energy transfer from BWDM to DR and constrain the dark-sector parameter space using the observational upper bound on the additional effective number of relativistic species, $\Delta N_{\mathrm{eff}}$.

The paper is organized as follows. We first introduce the microscopic interaction model and the phase-space distributions of the thermal and condensate components. We then derive the thermally averaged annihilation cross sections and investigate the relative importance of the three annihilation channels. Finally, we calculate the energy transfer to DR and discuss the resulting constraints on the BWDM parameter space from observational bounds on $\Delta N_{\mathrm{eff}}$.

We adopt the units $\hbar=c=k_{\rm B}=1$.

\section{Microscopic model}
\label{sec:Microscopic model}
\subsection{Interaction Lagrangian and annihilation process}
\label{sec:Interaction Lagrangian and annihilation process}

We consider a microscopic dark-sector model in which bosonic warm dark matter (BWDM) annihilates into relativistic dark radiation (DR). The Lagrangian density describing the BWDM field and the scalar DR field is given by \cite{Ludwick:2021}
\begin{equation}
\label{eq:Lagrangian density}
\mathcal{L}
=
\frac{1}{2}(\partial_{\mu}\chi)\partial^{\mu}\chi
-\frac{1}{2}M^2\chi^2
+
\frac{1}{2}(\partial_{\mu}\phi)\partial^{\mu}\phi
-\frac{1}{2}m^2\phi^2
-\frac{1}{2}g\phi\chi^2\,,
\end{equation}
where $\chi$ is a real scalar field representing the BWDM particle with mass $M$, while $\phi$ is a light real scalar field with mass $m$, whose relativistic excitations constitute the DR component. The trilinear coupling constant $g$ has mass dimension one, and the short-distance annihilation process is assumed to lie in the perturbative regime.

The trilinear interaction term $-g\phi\chi^2/2$ gives rise to the annihilation process
\begin{equation}
\label{eq:annihilation process}
\chi\chi\rightarrow\phi\phi\,.
\end{equation}
At tree level, this process proceeds through the $t$- and $u$-channel exchange of the $\chi$ field. The resulting annihilation transfers energy from the BWDM component to relativistic $\phi$ particles, providing a microscopic mechanism for DR production.

In addition, the exchange of the scalar field $\phi$ between non-relativistic BWDM particles induces an attractive interaction that can modify their two-particle wave function and lead to Sommerfeld enhancement of the annihilation rate. The derivation of the corresponding Yukawa potential and its implications for the annihilation process are discussed in Sec.~\ref{sec:Yukawa potential}.

\subsection{Annihilation cross section without Sommerfeld enhancement}
\label{sec:Tree-level annihilation cross section}

For simplicity, we consider the two-body annihilation process, $\chi_1\chi_2\rightarrow\phi_1\phi_2$.
Without the Sommerfeld enhancement induced by the long-range interaction, the tree-level annihilation cross section is given by
\begin{equation}
\label{equation:Cross sections without the interaction 1}
\sigma_{\rm free}
=
\frac{1}{\mathcal{S}}
\frac{1}{4E_{\chi_1}E_{\chi_2}\upsilon_{\chi}}
\left(
\prod_{j=1}^{2}
\int
\frac{{\rm d}^{3}k_{\phi_{\mathrm{j}}}}
{(2\pi)^3 2E_{\phi_{\mathrm{j}}}}
\right)
(2\pi)^4
\delta^{(4)}
\left(
k_{\chi_1}+k_{\chi_2}
-\sum_{j=1}^{2}k_{\phi_{\mathrm{j}}}
\right)
|\mathcal{M}|^2\,,
\end{equation}
where $E_{\chi_1}$ and $E_{\chi_2}$ are the energies of the initial BWDM particles, and $\upsilon_{\chi}$ denotes their relative velocity, $\upsilon_{\chi}=|\boldsymbol{\upsilon}_{\chi}|$. The factor $\mathcal{S}=2$ accounts for the identical particles in the final state.

The Lorentz invariant amplitude for the tree-level annihilation process is
\begin{equation}
\label{eq:the Lorentz invariant scattering amplitude}
|\mathcal{M}|^2
=
g^4
\left(
\frac{1}{M^2-t}
+
\frac{1}{M^2-u}
\right)^2\,,
\end{equation}
where $t=(k_{\chi_1}-k_{\phi_1})^2$, and $u=(k_{\chi_1}-k_{\phi_2})^2$ are the Mandelstam variables.

In the following, we consider the hierarchy $M\gg m$. Since the BWDM
particles are non-relativistic in the reference frame considered here, the center-of-mass frame of an annihilating BWDM pair is related to this frame by a non-relativistic boost. In the center-of-mass frame, energy conservation gives
\begin{equation}
E_{\phi}^*=E_{\chi}^* \geq M \gg m\,.
\end{equation}
Therefore,
\begin{equation}
k_{\phi}^*
=
\sqrt{(E_{\phi}^*)^2-m^2}
\simeq E_{\mathrm{\chi}}^*\gg m\,.
\end{equation}
Since the transformation back to the reference frame considered here
involves only a non-relativistic boost, the produced scalar particles
remain highly relativistic in this frame and therefore behave as DR.

Performing the phase-space integration in the center-of-mass frame, the tree-level annihilation cross section can be written as
\begin{equation}
\label{eq:computation result of cross sections without the interaction 1}
\sigma_{\rm free}
=
\frac{\sqrt{\frac{s-4m^2}{4s}}}
{64\pi E_{\chi_1}E_{\chi_2}\upsilon_{\chi}}
A(s)\,,
\end{equation}
where
\begin{equation}
A(s)
=
4g^4
\left[
\frac{1}{m^4-4m^2M^2+M^2s}
+
\frac{
4\operatorname{arctanh}
\left(
\frac{\sqrt{(s-4m^2)(s-4M^2)}}{2m^2-s}
\right)}
{(2m^2-s)
\sqrt{(s-4m^2)(s-4M^2)}}
\right]\,,
\end{equation}
and $s=(k_{\chi_1}+k_{\chi_2})^2$ is the Mandelstam variable. This result provides the perturbative annihilation cross section before including the non-perturbative Sommerfeld correction.

\subsection{The Yukawa potential in the non-relativistic limit}
\label{sec:Yukawa potential}
To calculate the Sommerfeld enhancement, we derive the effective
non-relativistic potential between two BWDM particles induced by the exchange of the DR scalar field $\phi$. The virtual exchange of this scalar field between BWDM particles generates the long-range Yukawa potential responsible for the Sommerfeld enhancement.

In the non-relativistic limit, the $s$-channel contribution does not generate a long-range interaction, since its propagator does not provide the momentum-transfer dependence required for a non-local potential. For identical BWDM particles, the $u$-channel contribution is related to the exchange symmetry of the two-body wave function. Although it contributes to the full scattering amplitude, it does not generate an independent long-range
potential kernel in the relative coordinate. Therefore, the long-range interaction relevant for the Sommerfeld enhancement is determined by the $t$-channel exchange of the DR scalar field $\phi$.

From the interaction term in Eq.\,\eqref{eq:Lagrangian density}, the
$t$-channel amplitude for the elastic BWDM scattering process, i.e.,
$\chi_1\chi_2\rightarrow\chi_1\chi_2$, is given by
\begin{equation}
\label{eq:scattering amplitude at t-channel}
\mathcal{M}_{\rm t}
=
-g^2\frac{1}{q^2-m^2+i\epsilon}\,,
\end{equation}
where $q=p_{\chi_1}-p'_{\chi_1}$ is the momentum transfer. Unlike the
annihilation process $\chi\chi\rightarrow\phi\phi$, the $t$-channel process involves a virtual exchange of the DR scalar field $\phi$, whose propagator generates the long-range interaction between BWDM particles.

In the non-relativistic limit, the energy transfer is much smaller than the spatial momentum transfer, i.e., $|q^0|\ll |\mathbf q|$. Therefore, the propagator denominator can be approximated as $q^2-m^2=(q^0)^2-\mathbf q^2-m^2\simeq-(\mathbf q^2+m^2)$.

The scattering amplitude becomes
\begin{equation}
\mathcal M'_t=
\frac{g^2}{|\mathbf q|^2+m^2}\,.
\end{equation}

The differential cross section in the non-relativistic limit is
\begin{equation}
\frac{d\sigma_t}{d\Omega}
=
\frac{|\mathcal M'_t|^2}{256\pi^2M^2}\,.
\end{equation}
On the other hand, in non-relativistic quantum mechanics, the Born scattering amplitude is related to the potential by
\begin{equation}
\mathcal{F}(\theta,\varphi)
=
-\frac{\mu_{\rm r}}{2\pi}
\int d^3R\,e^{i\mathbf q\cdot\mathbf R}V(\mathbf R),
\end{equation}
where $\mu_{\rm r}=M/2$ is the reduced mass. Matching the quantum mechanical result $\frac{d\sigma}{d\Omega}=|\mathcal{F}(\theta,\varphi)|^2$ with the field-theory expression yields the momentum-space potential
$V(\mathbf q)=-\frac{g^2}{4M^2}\frac{1}{\mathbf q^2+m^2}$. For a spherically symmetric potential, the Fourier transformation gives
\begin{equation}
V(R)
=
-\frac{g^2}{16\pi M^2}
\frac{e^{-mR}}{R}
=
-\frac{\alpha}{R}e^{-mR}\,,
\end{equation}
where $\alpha=\frac{g^2}{16\pi M^2}$. The above expression corresponds to the Yukawa potential generated by the
exchange of a massive scalar field \cite{Yukawa:1955}.

\subsection{Sommerfeld enhancement}
\label{sec:Sommerfeld enhancement}

After deriving the Yukawa potential in the non-relativistic limit, the Sommerfeld enhancement factor can be obtained by solving the corresponding two-body Schrödinger equation \cite{Arkani-Hamed:2009, Zavala:2009, Blum:2016, Coy:2022}. The Sommerfeld effect describes the modification of the annihilation probability due to the distortion of the two-particle wave function by the attractive long-range potential.

We consider the relative motion of two non-relativistic BWDM particles interacting through the potential $V(R)$. For an incoming plane wave propagating along the $z$-axis, the Schrödinger equation for the relative coordinate is given by
\begin{equation}
\label{eq:Schrödinger equation of the relative motion of two particles}
\left[
-\frac{\nabla^2}{2\mu_{\rm r}}
+
V(R)
\right]
\psi(\boldsymbol R)
=
E\psi(\boldsymbol R)\,,
\end{equation}
where $E=\frac12\mu_{\rm r}\upsilon_\chi^2$ is the kinetic energy associated with the relative motion, and
$\boldsymbol R=\boldsymbol R_1-\boldsymbol R_2$ is the relative coordinate between the two particles.

The Sommerfeld enhancement factor is determined by the ratio of the wave function at the origin with and without the long-range potential \cite{Cassel:2010, Coy:2022},
\begin{equation}
\label{eq:The Sommerfeld enhancement factor}
S=
\left|
\frac{\psi(0)}
{\psi_{\rm free}(0)}
\right|^2\,,
\end{equation}
where $\psi$ and $\psi_{\rm free}$ denote the solutions of Eq.\,\eqref{eq:Schrödinger equation of the relative motion of two particles} in the presence and absence of the Yukawa potential, respectively.

To obtain an analytic estimate of the Sommerfeld enhancement factor $S$, we introduce a characteristic length scale $R_{\rm c} \equiv 1/m$, which separates the short- and long-distance behavior of the interaction. At distances sufficiently smaller than $R_c$, the Yukawa potential approaches its Coulombic limit, whereas for $R\gtrsim R_c$ the interaction is exponentially suppressed. For an analytical estimate, we therefore approximate the potential by the piecewise form:
\begin{equation}\label{eq:approximated Yukawa potential}
V(R) \simeq
\begin{cases}
-\dfrac{\alpha}{R} & R < R_{\rm c}\,,\\[6pt]
0 & R \ge R_{\rm c}\,.
\end{cases}
\end{equation}
Although the Schrödinger equation with the exact Yukawa potential generally requires numerical solutions, the above approximation allows us to obtain an analytical estimate by using the Coulomb solution in the region where the wave-function distortion is dominated by the long-range attractive interaction \cite{Cassel:2010}.

Within this approximation, the Schrödinger equation in the region
$R<R_{\rm c}$ reduces to that for the attractive Coulomb potential.
The corresponding scattering solution is \cite{Gould:2006,Landau:1977}
\begin{equation}\label{eq:wave function with interaction}
\psi(R)=e^{\pi \beta / 2} \Gamma(1-i \beta) e^{i k z} F(i \beta ; 1 ; i k (R-z))\,,
\end{equation}
where $\beta=\frac{\alpha}{\upsilon_{\chi}}$ and $k=\mu_{\rm r}\upsilon_{\chi}$. The functions $\Gamma$ and $F$ are the
Gamma and confluent hypergeometric functions, respectively. When $\beta=0$, $\psi$ will return to $\psi_{\text{free}}$, that is,  
\begin{equation}\label{eq:wave function without interaction}
\psi_{\text{free}}(z)=e^{ikz}\,.    
\end{equation}
Substituting Eqs.\,\eqref{eq:wave function with interaction}-\eqref{eq:wave function without interaction} into Eq.\,\eqref{eq:The Sommerfeld enhancement factor}, the Sommerfeld enhancement factor is obtained as
\begin{equation}\label{eq:the result of the Sommerfeld enhancement factor}
S=\frac{2\pi\alpha/\upsilon_{\chi}}{1-e^{-2\pi\alpha/\upsilon_{\chi}}}\,.
\end{equation}
When the Yukawa interaction approaches the Coulomb limit, the Sommerfeld enhancement factor at low velocity behaves as
\begin{equation}
S\simeq\frac{2\pi\alpha}{v_\chi}\,.
\end{equation}
Therefore, in the Coulomb regime, the enhancement grows inversely with the relative velocity.

However, for a finite DR mass, the enhancement cannot increase indefinitely.
The Yukawa interaction has a finite range characterized by
$R_c=\frac1m$. The spatial variation of the two-body wave function is characterized by the de Broglie wavelength,
\begin{equation}
\lambda_{\rm dB}\sim\frac1{\mu_{\rm r} v_\chi}\,.
\end{equation}
When $\lambda_{\rm dB}\gtrsim R_c$, the wave function extends beyond the entire range of the Yukawa potential, and further decreasing the velocity does not lead to additional distortion of the wave function. Consequently, the Sommerfeld enhancement saturates at the velocity scale
\begin{equation}
v_{\rm sat}
\simeq
\frac1{\mu_{\rm r} R_c}
=
\frac{m}{\mu_{\rm r}}\,.
\end{equation}
Therefore, the velocity dependence of the Sommerfeld factor can be approximated as
\begin{equation}
S\simeq
\begin{cases}
\dfrac{2\pi\alpha/v_\chi}
{1-e^{-2\pi\alpha/v_\chi}},
&
v_\chi>v_{\rm sat},
\\[10pt]
S_{\rm sat},
&
v_\chi\leq v_{\rm sat},
\end{cases}
\end{equation}
where the saturation value can be estimated as
\begin{equation}
S_{\rm sat}\sim
\frac{2\pi\alpha}{v_{\rm sat}}
=
\frac{2\pi\alpha\mu_{\rm r}}{m}\,.
\end{equation}
Here, $S_{\mathrm{sat}}$ should be understood as an analytical estimate of the finite-range saturation of the Yukawa Sommerfeld enhancement.

\section{Phase-space distributions of BWDM}
We consider a single-species ideal Bose gas for BWDM, whose thermal component in kinetic equilibrium is described by the Bose--Einstein distribution \cite{zhang:2024},
\begin{equation}
\label{eq:HiDM-distribution}
f_{\chi_{\rm t}}(p;T,\mu)
=
\frac{1}{(2\pi)^3}
\frac{1}
{\exp[(\sqrt{p^2+M^2}-\mu)/T]-1}\,.
\end{equation}
Here, $T$ and $\mu$ denote the temperature and the chemical potential of the thermal component, respectively. Although the annihilation process $\chi\chi\rightarrow\phi\phi$ changes the BWDM particle number in principle, the interaction considered in this work is sufficiently weak. Therefore, the thermal component can be treated as approximately number conserving, and the Bose--Einstein distribution with a chemical potential is adopted to describe its phase-space distribution. We introduce the dimensionless variables
\begin{equation}
\label{eq:dimensionless parameter x}
x\equiv\frac MT\,,\qquad
x'\equiv\frac mT\,,
\end{equation}
and
\begin{equation}
\label{eq:dimensionless parameter delta}
\Delta\equiv\frac{\mu-M}{T}\,.
\end{equation}
Here, $x$ and $x'$ denote the ratios of the BWDM mass and the scalar DR mass to the temperature of the thermal component, respectively. The parameter $\Delta$ denotes the effective chemical potential-to-temperature ratio of the thermal component. The condition $\Delta=0$ corresponds to the critical point of Bose--Einstein condensation, where the chemical potential reaches $\mu=M$. The BWDM system is separated into a thermal component $\chi_{\rm t}$ and a condensate component $\chi_{\rm c}$. The condensate fraction is defined as
\begin{equation}
r\equiv\frac{n_{\chi_{\rm c}}}{n_{\chi_{\mathrm{tot}}}}\,,
\end{equation}
where
\begin{equation}
n_{\chi_{\mathrm{tot}}}
=
n_{\chi_{\rm t}}
+
n_{\chi_{\rm c}}\,,
\end{equation}
and $n_{\chi_{\mathrm{tot}}}$, $n_{\chi_{\rm t}}$, and $n_{\chi_{\rm c}}$ denote the total number density of BWDM particles, the number density of the thermal component, and the number density of the condensate component, respectively.

Assuming approximate conservation of the BWDM particle number, the condensate component is described as a macroscopic occupation of the zero-momentum state. In this work, we consider the evolution from the epoch of matter-radiation equality to the present epoch, during which BWDM particles have already entered the non-relativistic regime. If a condensate component has already existed at the beginning of matter-radiation equality, then under the conditions of entropy conservation and approximate particle-number conservation, the condensate fraction $r$ remains approximately constant during this non-relativistic evolution. Meanwhile, the existence of a condensate component requires the chemical potential of the thermal component to reach its critical value, corresponding to $\Delta=0$ in the Bose--Einstein distribution. The evolution of $r$ and $\Delta$ has been discussed in Ref.~\cite{zhang:2024}. Since the condensate particles occupy the zero-momentum state and their number density follows the dilution law of conserved particles, the phase-space distribution function of the condensate component is given by
\begin{equation}
\label{eq:distribution of BEC}
f_{\chi_{\rm c}}(\boldsymbol p)
=
\frac{r n_0}{a^3}\delta^{(3)}(\boldsymbol p)\,,
\end{equation}
where $n_0=\frac{\Omega_{\rm BWDM}\rho_{\mathrm{crit},0}}{M}$ is the present-day total BWDM number density, with $\Omega_{\rm BWDM}\simeq0.27$ denoting the BWDM density parameter and $\rho_{\mathrm{crit},0}\simeq3.74\times10^{-11}\,\mathrm{eV}^4$ denoting the present-day critical density of the Universe.

Assuming that the BWDM particles have no internal degrees of freedom, the total phase-space distribution is written as the sum of the thermal and condensate contributions,
\begin{equation}
f_{\chi_{\mathrm{tot}}}(\boldsymbol p,T)
=
f_{\chi_{\rm t}}(\boldsymbol p,T)
+
f_{\chi_{\rm c}}(\boldsymbol p)\,.
\end{equation}
The number density is obtained from the phase-space distribution as
\begin{equation}
n_{\chi}(T)
=
\int d^3p\,f_{\chi}(p,T)\,.
\end{equation}
The total energy density and pressure of the BWDM component are given by
\begin{equation}
\begin{aligned}
\rho_{\chi}(T)
&=
\int d^3p\,
f_{\chi}(p,T)E_{\chi}(p)\,,
\\
P_{\chi}(T)
&=
\int d^3p\,
f_{\chi}(p,T)
\frac{p^2}{3E_{\chi}(p)}\,,
\end{aligned}
\end{equation}
where $E_{\chi}(p)=\sqrt{p^2+M^2}$. Since the condensate component is located at zero momentum, its contribution to the pressure vanishes.

\section{Interaction between BWDM and DR}

\subsection{Thermally averaged annihilation cross section}

To calculate the energy transfer between BWDM and DR, we first evaluate the thermally averaged annihilation cross section including the Sommerfeld enhancement,
\begin{equation}
\label{eq:define of thermal average scattering cross-section}
\left\langle\sigma \upsilon_{\chi}\right\rangle
=
\frac{
\iint
\sigma\upsilon_{\chi}
f_{\chi}(\boldsymbol p_1)
f_{\chi}(\boldsymbol p_2)
\,d^3p_1d^3p_2
}
{
\iint
f_{\chi}(\boldsymbol p_1)
f_{\chi}(\boldsymbol p_2)
\,d^3p_1d^3p_2
}\,,
\end{equation}
where
\begin{equation}    
\sigma=\sigma_{\rm free}S
\end{equation}
is the annihilation cross section including the Sommerfeld enhancement factor.

In the presence of a BEC component, the initial BWDM states can be classified into three annihilation channels: the thermal–thermal (tt), thermal–condensate (ct), and condensate–condensate (cc) processes. Using the phase-space distributions introduced above and the dimensionless variables defined in
Eqs.\,\eqref{eq:dimensionless parameter x}--\eqref{eq:dimensionless parameter delta}, the thermally averaged annihilation cross sections in the non-relativistic regime considered here can be written as
\begin{equation}
\label{eq:the result for the three channels of thermal average scattering cross-section via the Sommerfeld correction}
\left\langle\sigma\upsilon_{\chi}\right\rangle
\simeq
\begin{cases}
\left\langle\sigma\upsilon_{\chi}\right\rangle_{\rm tt}
=
\dfrac{g^6}{1024\pi M^8}\,
\xi_{\rm tt}\,,
&
\mathrm{thermal–thermal},
\\[10pt]

\left\langle\sigma\upsilon_{\chi}\right\rangle_{\rm ct}
=
\dfrac{g^6}{512\pi M^8}\,
\xi_{\rm ct}\,,
&
\mathrm{thermal–condensate},
\\[10pt]

\left\langle\sigma\upsilon_{\chi}\right\rangle_{\rm cc}
=
\dfrac{g^6}{1024\pi M^7m}\,,
&
\mathrm{condensate–condensate}.
\end{cases}
\end{equation}
Here, $\xi_{\rm tt}$ and $\xi_{\rm ct}$ represent the dimensionless thermal averaging factors associated with the momentum distribution of the thermal component. Their detailed expressions are given in Appendix\,\ref{sec:The details of thermally averaged annihilation cross sections}.

For the condensate–condensate channel, both initial particles occupy the zero-momentum condensate state. Therefore, the relative velocity approaches zero, and the Sommerfeld enhancement reaches the saturation regime of the finite-range Yukawa potential. Consequently, the condensate–condensate annihilation cross section can be expressed analytically without additional thermal averaging.

The explicit derivations of the three annihilation channels, including the non-relativistic expansion of the Bose--Einstein distribution and the treatment of the Sommerfeld enhancement, are presented in Appendix\,\ref{sec:The details of thermally averaged annihilation cross sections}.

\subsection{Energy transfer from BWDM annihilation to dark radiation}\label{sec:The emergent energy-transfer term} With the above results for the thermally averaged annihilation cross sections, we now derive the effective dark-sector energy-transfer function induced by BWDM annihilation.

We only consider the annihilation process
$\chi\chi\rightarrow\phi\phi$ and neglect the inverse process $\phi\phi\rightarrow\chi\chi$.
We assume that the primordial abundance of the DR component is negligible, such that the DR particles considered here are produced through BWDM annihilation. In the weak-annihilation regime considered in this work, the resulting DR abundance remains small compared with the BWDM abundance, and hence the inverse reaction is strongly suppressed relative to the forward annihilation process. Therefore, its contribution to the evolution of the BWDM number density can be neglected. The Boltzmann equation describing the evolution of the BWDM number density \cite{Baumann2021Cosmology} is
\begin{equation}
\frac1{a^3}
\frac{d(n_{\chi_{\rm tot}}a^3)}{dt}
=
-n_{\chi_{\rm tot}}\Gamma_{\chi_{\rm tot}}\,,
\end{equation}
where
\begin{equation}\label{eq:the total interaction rate}  
\Gamma_{\chi_{\rm tot}}
=
\Gamma_{\chi_{\rm tt}}
+
2\Gamma_{\chi_{\rm ct}}
+
\Gamma_{\chi_{\rm cc}}\,,
\end{equation}

\begin{equation}\label{eq:the interaction rate between thermal component and thermal component}
\Gamma_{\chi_{\rm tt}}
=
(1-r)^2n_{\chi_{\rm tot}}
\langle\sigma v_\chi\rangle_{\rm tt}\,,
\end{equation}

\begin{equation}\label{eq:the interaction rate between thermal component and condensate component}
\Gamma_{\chi_{\rm ct}}
=
r(1-r)n_{\chi_{\rm tot}}
\langle\sigma v_\chi\rangle_{\rm ct}\,,
\end{equation}

\begin{equation}\label{eq:the interaction rate between condensate component and condensate component}
\Gamma_{\chi_{\rm cc}}
=
r^2n_{\chi_{\rm tot}}
\langle\sigma v_\chi\rangle_{\rm cc}\,.
\end{equation}
The factor of two in Eq.\,\eqref{eq:the total interaction rate} accounts for the two possible initial states,
$\chi_{\rm t}\chi_{\rm c}$ and $\chi_{\rm c}\chi_{\rm t}$.

During the evolution from the matter-dominated epoch to the present,
the BWDM particles are non-relativistic, and the BWDM component therefore satisfies $\rho_{\chi_{\rm tot}}\simeq Mn_{\chi_{\rm tot}}$
\cite{zhang:2024}. Using this relation, the above Boltzmann equation can be rewritten as
\begin{equation}\label{eq:the evolution of BWDM from Boltzmann equation}
\dot\rho_{\chi_{\rm tot}}
+
3H\rho_{\chi_{\rm tot}}
=
-\rho_{\chi_{\rm tot}}
\Gamma_{\chi_{\rm tot}}\,.
\end{equation}
Here, the Hubble parameter is taken as

\begin{equation}
H(a)
=
H_0
\sqrt{\Omega_m a^{-3}+\Omega_\Lambda}\,,
\end{equation}
where $\Omega_m\simeq0.31$ and $\Omega_\Lambda\simeq0.69$ are the present-day matter and cosmological-constant density parameters, respectively, and $H_0\simeq1.44\times10^{-33}\,\mathrm{eV}$ is the present-day Hubble constant.

On the other hand, the energy-transfer equations for the different components can be written separately. The BWDM component satisfies

\begin{equation}
\dot\rho_{\chi_{\rm tot}}
+
3H(\rho_{\chi_{\rm tot}}+p_{\chi_{\rm tot}})
=
Q\,.
\end{equation}

The DR satisfies the continuity equation
\begin{equation}\label{eq:the continuity equation of DR}
\dot\rho_{\phi}
+
3H(\rho_{\phi}+p_{\phi})
=
-Q\,.
\end{equation}

Since the BWDM pressure is negligible in the non-relativistic regime,
i.e., $p_{\chi_{\rm tot}}\ll\rho_{\chi_{\rm tot}}$, the BWDM continuity equation reduces to

\begin{equation}\label{eq:the evolution of BWDM from the continuity equation}
\dot\rho_{\chi_{\rm tot}}
+
3H\rho_{\chi_{\rm tot}}
=
Q\,.
\end{equation}

Comparing Eq.\,\eqref{eq:the evolution of BWDM from Boltzmann equation} and \eqref{eq:the evolution of BWDM from the continuity equation}, we obtain the effective energy-transfer term generated by BWDM annihilation,

\begin{equation}
\boxed{
Q
=
-\rho_{\chi_{\rm tot}}
\Gamma_{\chi_{\rm tot}}
}\,.
\end{equation}
We adopt the convention that a negative $Q$ corresponds to energy transfer from BWDM to the DR.

\begin{figure}[htbp]
\centering
\includegraphics[width=1.0\textwidth]{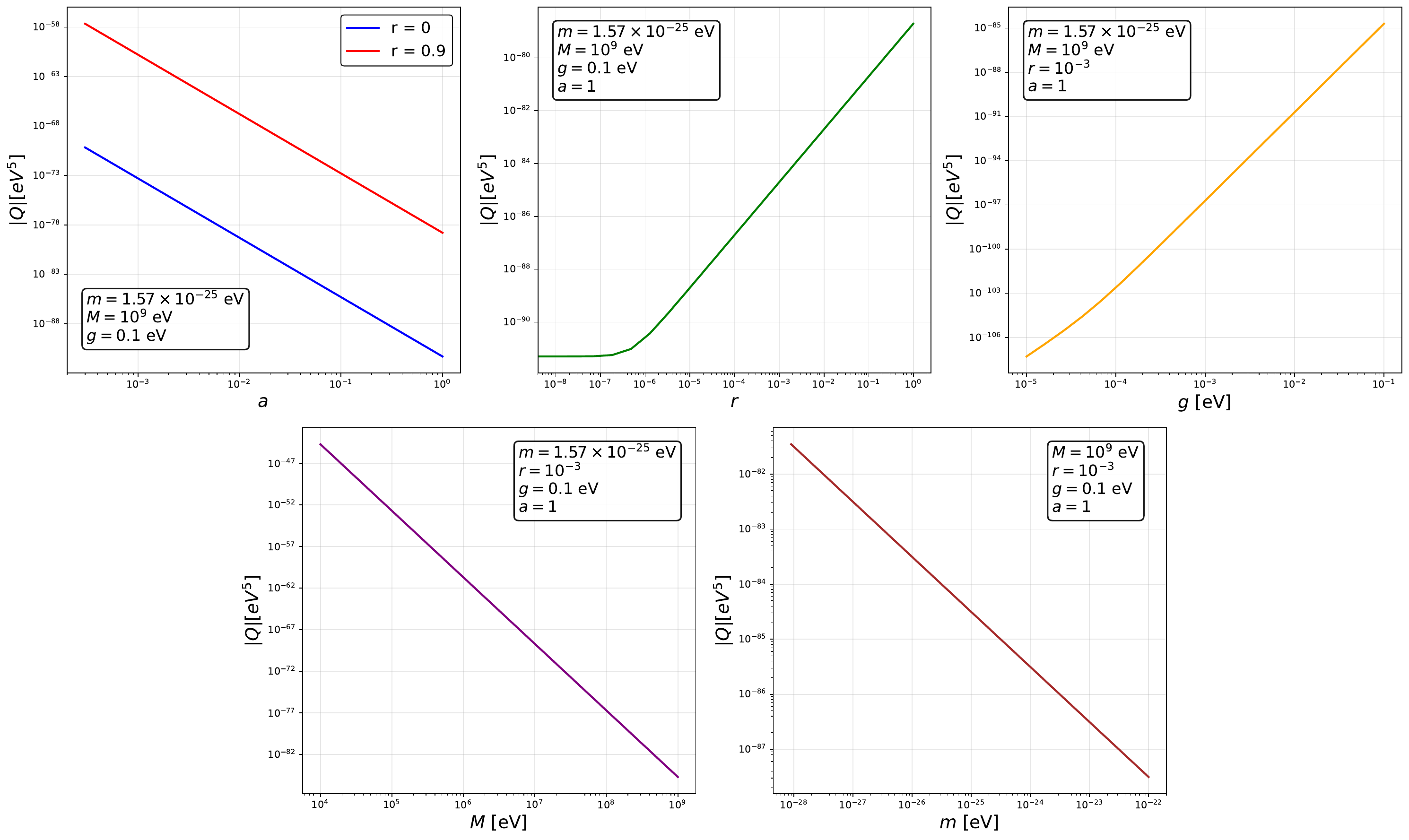}
\qquad
\caption{The dependence of the energy-transfer term $|Q|$ on the cosmological scale factor $a$, the condensate fraction $r$, the BWDM mass $M$, the coupling constant $g$, and the scalar DR mass $m$. The scalar DR mass is fixed to $m=1.57\times10^{-25}\,\mathrm{eV}$ when the dependence on other parameters is investigated. The upper-left panel compares the evolution of $|Q|$ for a pure thermal BWDM scenario ($r=0$) and a mixed thermal--condensate scenario ($r=0.9$), with $M=10^9\,\mathrm{eV}$ and $g=0.1\,\mathrm{eV}$. The remaining panels show the dependence on $r$, $g$, $M$, and $m$ with the other parameters fixed as indicated in each panel. All results are obtained by numerically evaluating the full thermally averaged annihilation cross sections.
\label{fig:Q_dependence}}
\end{figure}

The above relation establishes a direct connection between the microscopic annihilation processes and the macroscopic BWDM--DR energy exchange. By substituting the contributions from the three annihilation channels, together with the thermally averaged annihilation cross sections obtained above, the energy-transfer term can be determined as a function of the fundamental parameters,
\begin{equation}
Q=Q(a,r,M,m,g)\,.
\end{equation}
Since the energy transfer is always from BWDM to DR in the scenario considered here, we focus on the magnitude of the energy-transfer term, $|Q|$, in the following discussion. The dependence of $|Q|$ on several fundamental parameters is shown in Figure\,\ref{fig:Q_dependence}. The results are obtained by numerically evaluating the full thermally averaged annihilation cross sections given in Eq.\,\eqref{eq:full thermal averaged annihilation cross sections}, where the velocity dependence of the Sommerfeld enhancement and the momentum distributions are included.

The upper-left panel of Figure\,\ref{fig:Q_dependence} presents the evolution of $|Q|$ with the scale factor $a$. As the universe expands, the magnitude of the energy-transfer term decreases mainly due to the cosmological dilution of the BWDM number density. In the non-relativistic regime considered here, the BWDM temperature is sufficiently low that the thermally averaged annihilation cross sections without Sommerfeld enhancement approach approximately constant values \cite{Weinberg:2008}. The Sommerfeld enhancement modifies the annihilation rates through the relative velocity of the annihilating particles, but does not introduce an additional explicit dependence on the scale factor in the parameter region considered here. Therefore, the evolution of $|Q|$ is predominantly governed by the decrease of the BWDM number densities with cosmic expansion. Nevertheless, the presence of a BEC component significantly enhances the interaction compared with the pure thermal BWDM case, mainly through the condensate--condensate contribution.

The upper-middle panel of Figure\,\ref{fig:Q_dependence} illustrates the dependence of $|Q|$ on the condensate fraction $r$. The variation of the energy-transfer term with $r$ reflects the modification of the annihilation processes caused by the presence of the condensate component. As the condensate fraction increases, the contribution from the BEC-related channels becomes increasingly important due to the distinct annihilation properties of the condensate particles. Although the condensate--condensate contribution is suppressed by the condensate abundance factor $r^2$, its annihilation cross section can be significantly enhanced because the condensate particles occupy the zero-momentum ground state. Consequently, the condensate--condensate channel probes the saturated low-velocity limit of the Sommerfeld enhancement induced by the finite-range Yukawa interaction, leading to a substantial increase of the corresponding interaction contribution. Therefore, the presence of a BEC component provides an efficient mechanism for modifying the effective BWDM-to-DR energy-transfer rate. The transition between the thermal- and condensate-dominated annihilation regimes is analyzed in detail in Sec.\,\ref{sec:Dominant annihilation channels and the critical condensate fraction}.

The upper-right panel of Figure\,\ref{fig:Q_dependence} presents the dependence of $|Q|$ on the coupling constant $g$. As expected, increasing the coupling strength enhances the interaction rate. This dependence originates from both the tree-level annihilation amplitude and the Sommerfeld enhancement factor, which are strengthened by a larger coupling. Consequently, a stronger BWDM--DR coupling leads to a more efficient energy transfer from BWDM to DR.

The lower-left panel of Figure\,\ref{fig:Q_dependence} shows the dependence of $|Q|$ on the BWDM mass $M$. We find that the energy-transfer term decreases rapidly with increasing $M$. This behavior results from the combined effects of the mass dependence of the BWDM number density and the strong dependence of the thermally averaged annihilation cross sections on the BWDM mass. Therefore, heavier BWDM particles generally lead to a suppressed late-time energy transfer between BWDM and DR within the parameter space considered here.

The lower-right panel of Figure\,\ref{fig:Q_dependence} presents the dependence of $|Q|$ on the scalar DR mass $m$. As the scalar DR mass increases, the range of the Yukawa interaction becomes shorter, which suppresses the Sommerfeld enhancement. Consequently, the BWDM annihilation rate and the corresponding energy-transfer term decrease for larger scalar DR masses.

\section{Dominant annihilation channels and the critical condensate fraction}
\label{sec:Dominant annihilation channels and the critical condensate fraction}
The total energy-transfer term $Q$ receives contributions from three annihilation channels, including the thermal--thermal (tt), thermal--condensate (tc), and condensate--condensate (cc) processes. Although all three channels are included in the numerical calculation, their relative contributions can be significantly different due to the distinct phase-space distributions and Sommerfeld enhancements of the thermal and condensate components.

Since the BWDM particles have already entered the non-relativistic regime during the epoch considered in this work, the thermally averaged annihilation cross sections remain approximately constant. The dominant scale-factor dependence of the interaction rates originates from the cosmological dilution of the BWDM number densities. When comparing the relative importance of different annihilation channels, the common dilution factor cancels in the ratios of the interaction rates. Therefore, the hierarchy among the annihilation channels can be evaluated using the present-day interaction rates at $a=1$.

\begin{figure}[htbp]
\centering

\includegraphics[width=0.65\textwidth]{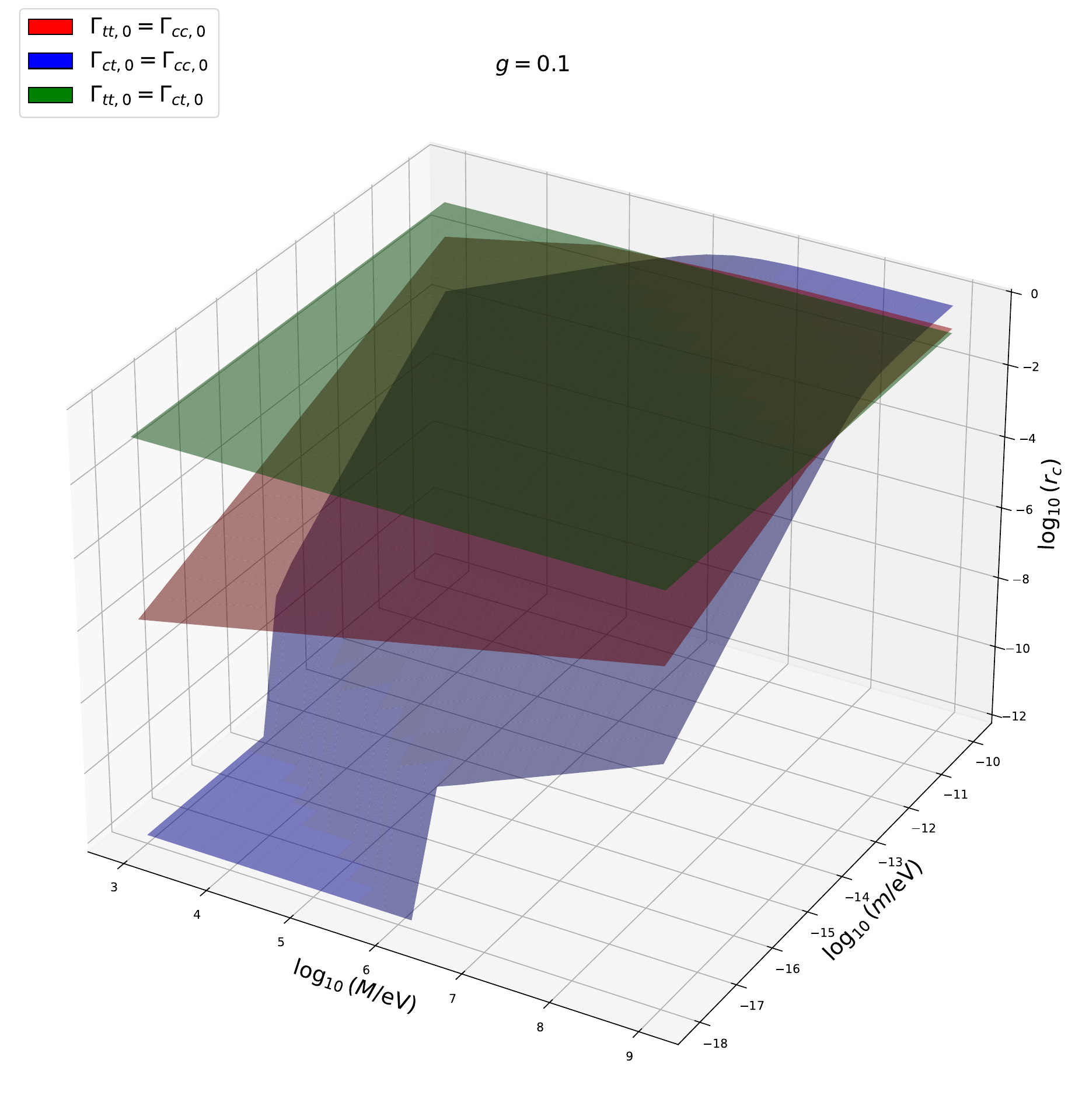}

\vspace{0.3cm}

\includegraphics[width=0.65\textwidth]{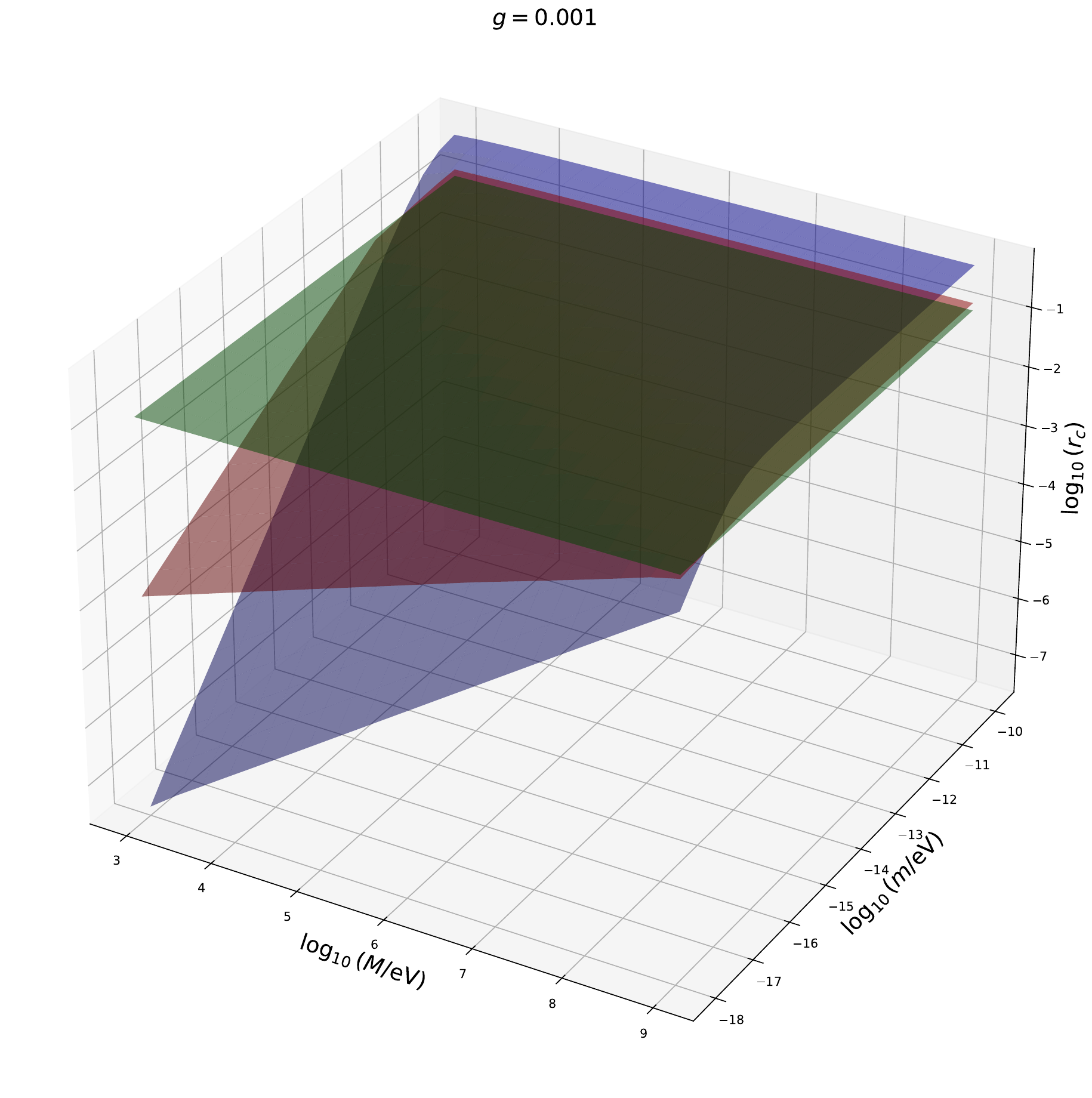}

\caption{
Critical surfaces corresponding to
$\Gamma_{\chi_{\mathrm{tt, 0}}}=\Gamma_{\chi_\mathrm{cc, 0}}$
(red),
$\Gamma_{\chi_{\mathrm{ct, 0}}}=\Gamma_{\chi_\mathrm{cc, 0}}$
(blue), and
$\Gamma_{\chi_{\mathrm{tt, 0}}}=\Gamma_{\chi_{\mathrm{ct, 0}}}$
(green).
The upper and lower panels correspond to
$g=0.1$ and $g=0.001$, respectively.
The comparison shows that the thermal--condensate channel remains subdominant in the parameter region considered.
}
\label{fig:critical surfaces}
\end{figure}

To determine the dominant annihilation channels, we compare the interaction rates of the three contributions by introducing the critical surfaces defined by
\begin{equation}
\Gamma_{\chi_{\mathrm{tt, 0}}}=\Gamma_{\chi_\mathrm{cc, 0}}\,,
\qquad
\Gamma_{\chi_{\mathrm{ct, 0}}}=\Gamma_{\chi_\mathrm{cc, 0}}\,,
\qquad
\Gamma_{\chi_{\mathrm{tt, 0}}}=\Gamma_{\chi_{\mathrm{ct, 0}}}\,,
\end{equation}
where the subscript $0$ denotes the present-day value. The corresponding critical surfaces are shown in Fig.\,\ref{fig:critical surfaces}. These results demonstrate that, for the relatively small scalar DR masses considered in this work, the thermal--condensate contribution remains subdominant compared with the other two channels. Therefore, the transition of the dominant annihilation contribution is mainly controlled by the competition between the thermal--thermal and condensate--condensate channels.

To quantify the hierarchy between the thermal--thermal and condensate--condensate channels, we compare their thermally averaged annihilation cross sections. Using Eq.\,\eqref{eq:analytical thermal averaged cross sections appendix}, we obtain
\begin{equation}
\frac{
\left\langle\sigma\upsilon_{\chi}\right\rangle_{\rm tt}
}{
\left\langle\sigma\upsilon_{\chi}\right\rangle_{\rm cc}
}
\sim
\frac{m}{M}
\sqrt{\frac{x}{2}}
\sim
1.87\times10^{3}a
\left(\frac{m}{\mathrm{eV}}\right)
\left(
\frac{M/\mathrm{eV}}{1-r}
\right)^{1/3}\,.
\end{equation}
Cosmological constraints on strong dark forces with interaction ranges $R_{\rm c}\lesssim100\,\mathrm{kpc}$ have been investigated in Ref.\,\cite{Graham:2026}. Motivated by this characteristic length scale, we restrict the range of the Yukawa interaction in our analysis to $R_{\rm c}\lesssim100\,\mathrm{kpc}$. Since the interaction range is determined by the mass of the scalar DR field through $R_{\rm c}\sim m^{-1}$, this condition corresponds to
\begin{equation}
m\gtrsim6.4\times10^{-29}\,\mathrm{eV}\,.
\end{equation}
To estimate the parameter region where the hierarchy between the thermal--thermal and condensate--condensate annihilation cross sections could become less pronounced, we consider the condition
\begin{equation}
\frac{
\left\langle\sigma\upsilon_{\chi}\right\rangle_{\rm tt}
}{
\left\langle\sigma\upsilon_{\chi}\right\rangle_{\rm cc}
}
\sim
\frac{1}{(1-r)^{1/3}}\,.
\end{equation}
This condition corresponds to
\begin{equation}
M
\gtrsim
\frac{1}{(1.84\times10^3a\,m/\mathrm{eV})^3}\,
\mathrm{eV}\,.
\end{equation}
For simplicity, we consider the scalar DR mass range $6.4\times10^{-29}\,\mathrm{eV}\lesssim m\lesssim10^{-18}\,\mathrm{eV}$. At the present epoch ($a=1$), the smallest BWDM mass threshold within this range is obtained for $m=10^{-18}\,\mathrm{eV}$, yielding
\begin{equation}
M\sim10^{44}\,\mathrm{eV}\,,
\end{equation}
which is far beyond any viable particle DM mass scale. For smaller scalar DR masses within the considered interval, the corresponding BWDM mass threshold becomes even larger. Consequently, within the physically relevant parameter space considered here, we have
\begin{equation}
\frac{
\left\langle\sigma\upsilon_{\chi}\right\rangle_{\rm tt}
}{
\left\langle\sigma\upsilon_{\chi}\right\rangle_{\rm cc}
}
\ll
\frac{1}{(1-r)^{1/3}}\,.
\end{equation}
This hierarchy implies that the strong Sommerfeld enhancement of the zero-momentum condensate particles can compensate for the suppression of the condensate number density, allowing the condensate--condensate channel to become comparable to the thermal--thermal channel even for a relatively small condensate fraction.

The critical condensate fraction $r_{\rm c}$ is determined by the equality of the thermal--thermal and condensate--condensate interaction rates,
\begin{equation}
\Gamma_{\chi_{\rm tt}}
=
\Gamma_{\chi_{\rm cc}} .
\label{eq:definition rc}
\end{equation}
Using Eq.\,\eqref{eq:the interaction rate between thermal component and thermal component} and
Eq.\,\eqref{eq:the interaction rate between condensate component and condensate component}, we obtain

\begin{equation}
\left(
\frac{r_{\rm c}}{1-r_{\rm c}}
\right)^2
=
\frac{
\left\langle\sigma\upsilon_{\chi}\right\rangle_{\mathrm{tt}}
}
{
\left\langle\sigma\upsilon_{\chi}\right\rangle_{\mathrm{cc}}
}.
\end{equation}
Since the right-hand side is strongly suppressed,
$r_{\rm c}$ satisfies

\begin{equation}
r_{\rm c}\ll1 .
\end{equation}
Therefore, $(1-r_{\rm c})^2$ can be approximated as unity, leading to

\begin{equation}
\label{eq:analytical expression of rc}
r_{\rm c}
\simeq
\left(
\frac{
\left\langle\sigma\upsilon_{\chi}\right\rangle_{\mathrm{tt}}
}
{
\left\langle\sigma\upsilon_{\chi}\right\rangle_{\mathrm{cc}}
}
\right)^{1/2}.
\end{equation}
Substituting the analytical approximations of the thermally averaged annihilation cross sections given in Eq.\,\eqref{eq:analytical thermal averaged cross sections appendix}, we obtain an analytical estimate of the dependence of $r_{\rm c}$ on the microscopic parameters.

Our numerical results show that the critical condensate fraction $r_{\rm c}$ remains nearly constant throughout the evolution from the matter-dominated epoch to the present. This result is qualitatively consistent with the slow evolution of the condensate fraction under particle-number and specific-entropy conservation discussed in Ref.\,\cite{zhang:2024}.

\begin{figure}[htbp]
\centering
\includegraphics[width=1.0\textwidth]{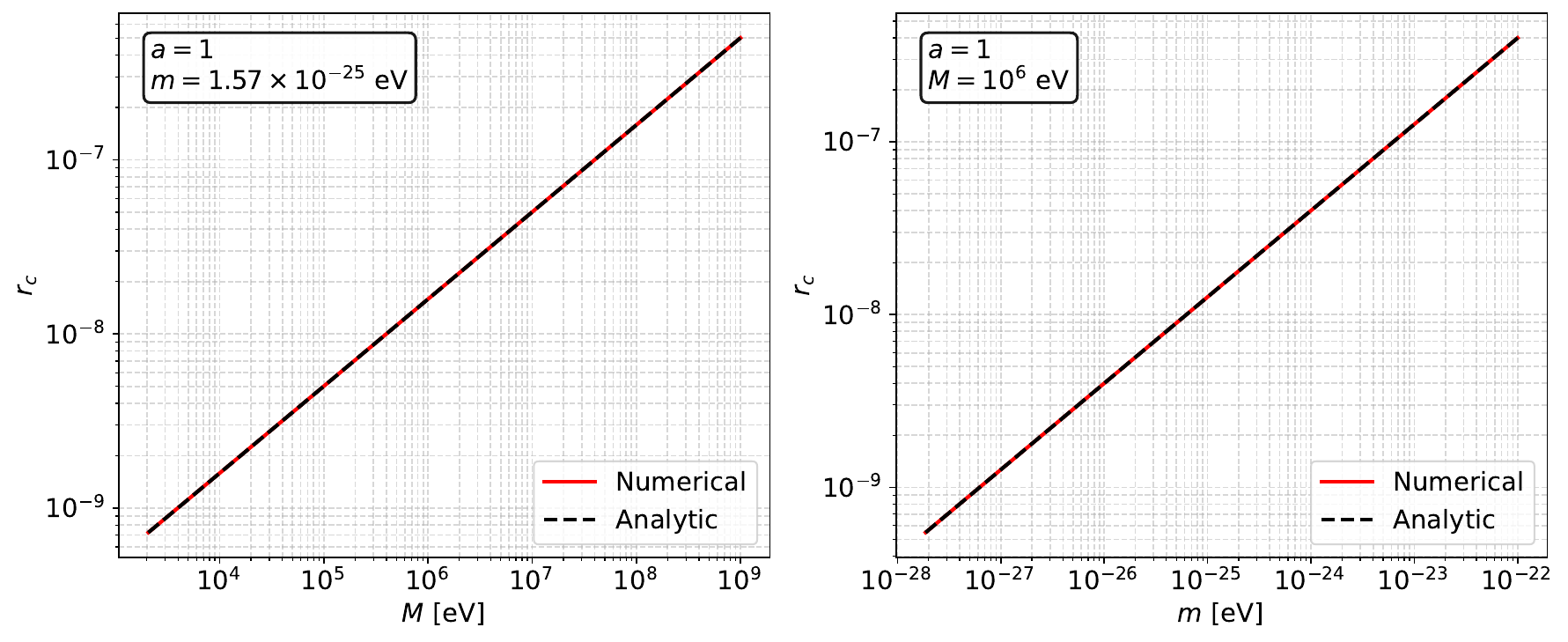}
\caption{The critical condensate fraction $r_{\rm c}$ determined from
$\Gamma_{\chi_{\rm tt}}=\Gamma_{\chi_{\rm cc}}$. The red solid curves represent the numerical solutions obtained using Eq.\,\eqref{eq:full thermal averaged annihilation cross sections} given in Appendix\,\ref{sec:The details of thermally averaged annihilation cross sections}.
The black dashed curves represent the analytical approximation obtained from
Eq.\,\eqref{eq:analytical expression of rc} and Eq.\,\eqref{eq:analytical thermal averaged cross sections appendix}. The left and right panels show the dependence of $r_{\rm c}$ on the BWDM mass $M$ and the scalar DR mass $m$, respectively, at the present epoch.}
\label{fig:the variation of rc with M or m}
\end{figure}

The numerical solution of Eq.\,\eqref{eq:definition rc} and the corresponding analytical approximation are shown in Fig.\,\ref{fig:the variation of rc with M or m}.

As shown in Fig.\,\ref{fig:the variation of rc with M or m}, the analytical approximation agrees well with the numerical solution over the parameter region considered. This agreement verifies the non-relativistic expansion used in obtaining the analytical cross sections and the approximation $r_{\rm c}\ll1$.

For fixed scalar DR mass $m$, $r_{\rm c}$ increases with increasing BWDM mass $M$. This behavior originates from the different mass dependence of the thermal--thermal and condensate--condensate annihilation channels.

For fixed BWDM mass $M$, increasing the scalar DR mass also increases $r_{\rm c}$. A heavier scalar DR particle shortens the interaction range of the Yukawa potential and weakens the Sommerfeld enhancement, thereby reducing the relative contribution of the condensate--condensate channel. Consequently, a larger condensate fraction is required for the condensate--condensate channel to become comparable to the thermal--thermal channel.

Therefore, the critical condensate fraction separates two different interaction regimes,

\begin{equation}
\begin{cases}
r<r_{\rm c},
&
\Gamma_{\chi_{\mathrm{tt}}}>
\Gamma_{\chi_{\mathrm{cc}}},
\\[6pt]
r>r_{\rm c},
&
\Gamma_{\chi_{\mathrm{cc}}}>
\Gamma_{\chi_{\mathrm{tt}}}.
\end{cases}
\end{equation}

The corresponding constraints on the BWDM mass from the CMB observations are discussed in Sec.\,\ref{sec:Constraints on the dark-sector parameter space from CMB}.

\section{Constraints on the dark-sector parameter space from $\Delta N_{\mathrm{eff}}$}
\label{sec:Constraints on the dark-sector parameter space from CMB}
The critical condensate fraction obtained in Sec.\,\ref{sec:Dominant annihilation channels and the critical condensate fraction} determines the transition between different annihilation-dominated regimes. In this section, we investigate how this transition affects the constraints on the BWDM parameter space derived from observational bounds on $\Delta N_{\mathrm{eff}}$.

The annihilation of BWDM particles into scalar DR contributes to the additional radiation energy density, which can be constrained through the effective number of relativistic species, $N_{\mathrm{eff}}$. As derived in Appendix~\ref{sec:The relation between Delta N eff and the current rate of interaction}, the observational upper bound on $\Delta N_{\mathrm{eff}}$ leads to the following constraint on the present-day BWDM annihilation rate:
\begin{equation}\label{eq:the CMB constraint on the BWDM annihilation rate}
\Gamma_{\chi_{\mathrm{tot}},0}
\leq
\frac{\Omega_{\gamma}H_0\sqrt{\Omega_m a_{\mathrm{eq}}}}
{8.8\Omega_{\mathrm{BWDM}}}
\Delta N_{\mathrm{eff}}^{\mathrm{max}}\,,
\end{equation}
where $\Delta N_{\mathrm{eff}}^{\mathrm{max}}=0.107$ is the 95\% confidence-level upper bound obtained from the combined primordial-abundance, CMB, and baryon acoustic oscillation (BAO) measurements \cite{Goldstein:2026}. This constraint relates the microscopic parameters of the dark sector, including the BWDM mass $M$, the scalar DR mass $m$, the coupling constant $g$, and the condensate fraction $r$, to the observational bound on the additional radiation density.

For fixed $(m,g,r)$, the critical BWDM mass $M_{\rm c}$ is determined by saturating this bound:
\begin{equation}\label{eq:the equation of critical BWDM mass}
\Gamma_{\chi_{\mathrm{tot}},0}(M_{\rm c},m,g,r)
=
\frac{\Omega_{\gamma}H_0\sqrt{\Omega_m a_{\mathrm{eq}}}}
{8.8\Omega_{\mathrm{BWDM}}}
\Delta N_{\mathrm{eff}}^{\mathrm{max}}\,.
\end{equation}
In the parameter region considered here, the annihilation rate decreases with increasing BWDM mass. Therefore, the parameter region satisfying the observational bound can be expressed as
\begin{equation}
M\geq M_{\rm c}\,.
\end{equation}

The red solid line in the left panel of
Fig.\,\ref{fig:the critical mass varies with r, m and g}
is obtained by numerically solving
Eq.\,\eqref{eq:the equation of critical BWDM mass}, where the full thermally averaged annihilation cross sections given in Eq.\,\eqref{eq:full thermal averaged annihilation cross sections}
are adopted.

\begin{figure}[htbp]
\centering
\includegraphics[width=1.0\textwidth]{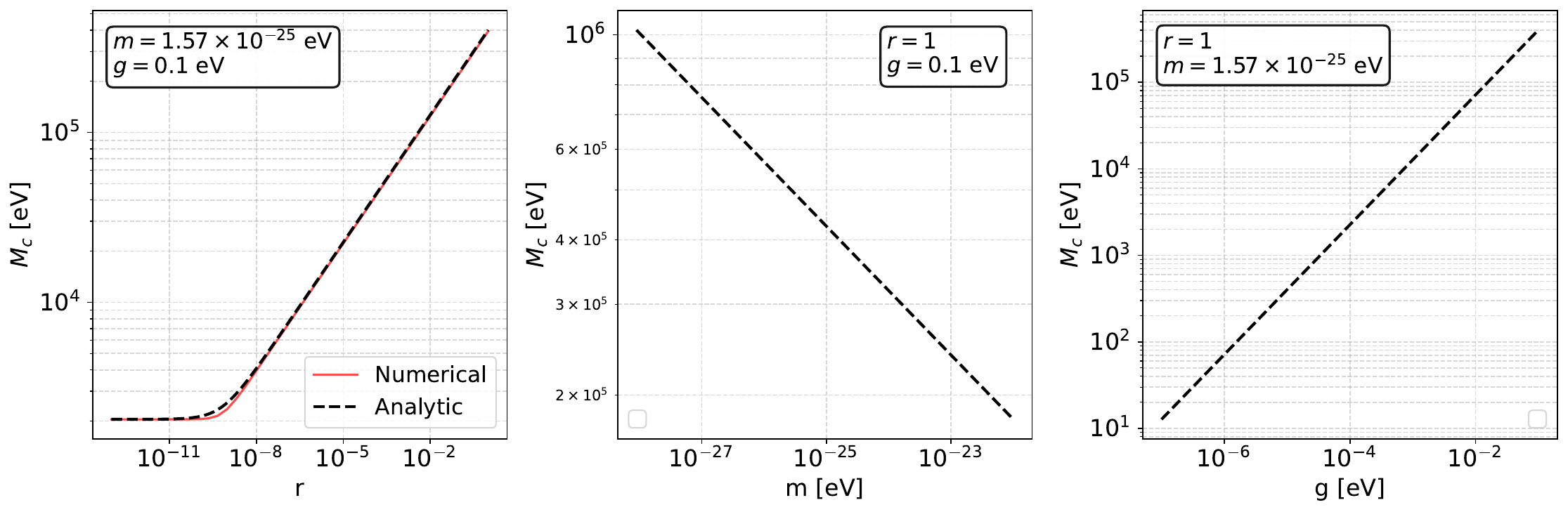}
\qquad
\caption{
Constraints on the dark-sector parameter space represented by the critical BWDM mass $M_c$, obtained from the observational upper bound on $\Delta N_{\mathrm{eff}}$ through Eq.\,\eqref{eq:the CMB constraint on the BWDM annihilation rate}.
The left panel shows the dependence of $M_c$ on the condensate fraction $r$
for a general mixed thermal--condensate state. The red solid curve is obtained by numerically solving Eq.\,\eqref{eq:the equation of critical BWDM mass}, where the thermally averaged annihilation cross sections are evaluated using the full expressions in Eq.\,\eqref{eq:full thermal averaged annihilation cross sections}.
The black dashed curve corresponds to the analytical approximation obtained from the simplified thermally averaged annihilation cross sections. The middle and right panels show the dependence of $M_c$ on the scalar DR mass $m$ and the coupling constant $g$, respectively, in the pure BEC limit ($r=1$). The critical mass in these two panels is obtained from Eq.\,\eqref{eq:critical mass above rc}. The fixed parameters are $m=1.57\times10^{-25}\,\mathrm{eV}$ and $g=0.1\,\mathrm{eV}$ for the left panel, $g=0.1\,\mathrm{eV}$ for the middle panel, and $m=1.57\times10^{-25}\,\mathrm{eV}$ for the right panel.
}
\label{fig:the critical mass varies with r, m and g}
\end{figure}

To understand the dependence of the critical mass on the condensate fraction, we analyze two limiting regimes according to the dominant annihilation channels discussed in
Sec.\,\ref{sec:Dominant annihilation channels and the critical condensate fraction}.

For a small condensate fraction, $r<r_{\rm c}$, the thermal–thermal channel dominates the total interaction rate. Therefore, the interaction rate can be approximated as
\begin{equation}
\Gamma_{\chi_{\mathrm{tot}},0}
\simeq
\Gamma_{\chi_{\mathrm{tt}},0}
=
(1-r)^2
n_{\chi_{\mathrm{tot}},0}
\langle\sigma\upsilon_\chi\rangle_{\mathrm{tt, 0}}\,.
\end{equation}
Since the critical condensate fraction satisfies $r_{\rm c}\ll1$, the factor $(1-r)^2$ remains close to unity in this regime.
Consequently, the critical mass approaches the value determined by the thermal–thermal interaction,
\begin{equation}
M_{\rm c}\simeq2.048\times10^{3}\,\mathrm{eV}.
\end{equation}

For a sufficiently large condensate fraction, i.e., $r>r_{\rm c}$,
the condensate–condensate channel becomes the dominant contribution, and the interaction rate is approximately given by
\begin{equation}
\Gamma_{\chi_{\mathrm{tot}},0}
\simeq
\Gamma_{\chi_{\mathrm{cc}},0}
=
r^2
n_{\chi_{\mathrm{tot}},0}
\langle\sigma\upsilon_\chi\rangle_{\rm cc}\,.
\end{equation}
Using the analytical expression for the condensate--condensate annihilation cross section, the CMB constraint leads to
\begin{equation}
\label{eq:critical mass above rc}
M_{\rm c}
=
4.02\times10^5
r^{0.25}
\left(
\frac{g}{g_{\mathrm{ref}}}
\right)^{0.75}
\left(
\frac{m}{m_{\mathrm{ref}}}
\right)^{-0.125}
\mathrm{eV}\,,
\end{equation}
where $g_{\mathrm{ref}}=0.1\,\mathrm{eV}$ and $m_{\mathrm{ref}}=1.57\times10^{-25}\,\mathrm{eV}$.

The above two limiting behaviors provide an analytical understanding of the condensate-fraction dependence of the critical mass. Motivated by these two regimes, we introduce an effective expression,
\begin{equation}
\label{eq:effective critical mass parametrization}
M_{\rm c}
=
4.39\times10^5
(r+c)^{0.25}
\left(
\frac{g}{g_{\mathrm{ref}}}
\right)^{0.75}
\left(
\frac{m}{m_{\mathrm{ref}}}
\right)^{-0.125}
\mathrm{eV}\,.
\end{equation}
Here, the parameter $c$ is determined by matching the thermal-dominated limit and the BEC-dominated limit. Using the critical mass obtained from the thermal–thermal channel together with the analytical condensate–condensate expression, we obtain $c=6.736\times10^{-10}$.

Therefore, $c$ is not an empirical fitting parameter, but represents the characteristic condensate fraction associated with the transition between the two annihilation regimes.

For $r\ll c\simeq r_{\rm c}$, the condensate contribution is insufficient to modify the interaction rate significantly, and the critical mass approaches the thermal BWDM result. In contrast, for $r\gg c\simeq r_{\rm c}$, the condensate–condensate channel dominates due to the enhanced annihilation rate of the zero-momentum condensate component, and the scaling behavior approaches the analytical condensate–condensate result given by Eq.\,\eqref{eq:critical mass above rc}.

The black dashed curve in the left panel of
Fig.\,\ref{fig:the critical mass varies with r, m and g}
is obtained from Eq.\,\eqref{eq:effective critical mass parametrization}. The excellent agreement between the analytical parametrization and the full numerical result demonstrates that the two limiting annihilation regimes capture the dominant dependence of the critical mass on the condensate fraction.

The middle and right panels of Fig.\,\ref{fig:the critical mass varies with r, m and g} are obtained from the BEC dominated analytical relation, Eq.\,\eqref{eq:critical mass above rc}, corresponding to the pure condensate limit.

The middle panel shows the dependence of the critical mass on the scalar DR mass $m$. A smaller DR mass corresponds to a longer interaction range and a stronger Sommerfeld enhancement, resulting in a larger annihilation rate. Therefore, a larger BWDM mass is required to satisfy the CMB constraint on $\Delta N_{\mathrm{eff}}$.

The right panel presents the dependence of the critical mass on the coupling strength $g$. Since the annihilation rate increases with increasing coupling strength, larger values of $g$ require a larger BWDM mass to satisfy the CMB constraint on $\Delta N_{\mathrm{eff}}$.

\section{Discussion}

The dark-sector energy transfer considered here originates from the microscopic annihilation process $\chi\chi\rightarrow\phi\phi$, rather than from a phenomenological parameterization of $Q$. Its magnitude is therefore determined by the particle-physics parameters and the BWDM phase-space distribution. The relativistic scalar particles produced in the annihilation constitute the DR component, providing a microscopic mechanism for energy transfer from BWDM to DR.

A distinctive feature of our framework is that the BWDM component is assumed to contain a Bose--Einstein condensed (BEC) component. By decomposing the BWDM phase-space distribution into thermal and condensate components, the annihilation rate naturally receives contributions from three channels: thermal--thermal, thermal--condensate, and condensate--condensate annihilation. These channels have different phase-space distributions and therefore contribute differently to the thermally averaged annihilation rates. In particular, the condensate--condensate channel probes the zero-relative-velocity limit because both annihilating particles occupy the zero-momentum condensate state. For the finite-range Yukawa interaction considered here, this channel consequently probes the saturated low-velocity regime of the Sommerfeld enhancement. Although low-velocity particle pairs also occur in the thermal channels, they constitute only a fraction of the thermal momentum distributions. The condensate--condensate channel can therefore receive a substantially larger Sommerfeld enhancement than the thermal channels in the parameter region considered here.

The main physical effect studied in this work is not merely the Sommerfeld enhancement itself. Rather, the BEC fraction provides an additional quantum-statistical degree of freedom that controls the relative importance of the microscopic annihilation channels and, consequently, the efficiency of energy transfer from BWDM to DR. The competition between the thermal--thermal and condensate--condensate channels determines the transition between different annihilation-dominated regimes, characterized by the critical condensate fraction $r_{\rm c}$. Our numerical results indicate that $r_{\rm c}$ remains nearly constant over the non-relativistic cosmological evolution considered here. Although the thermal--condensate contribution is explicitly included in the calculation, it remains subdominant for the relatively small scalar DR masses considered in this work. The critical condensate fraction therefore provides a useful characterization of the transition between the thermal--thermal and condensate--condensate dominated regimes.

In this work, we assume that a highly condensed BWDM component is present in the early Universe and investigate its implications for the annihilation processes and the resulting CMB constraints. Our analysis focuses on the non-relativistic regime, in which the condensate fraction $r$ remains approximately constant under particle-number and specific-entropy conservation, and a sufficiently high condensate fraction at the onset of the non-relativistic regime can persist up to the present epoch \cite{zhang:2024}. The dynamical mechanism responsible for generating such a high condensate fraction is beyond the scope of the present work.

We further use the observational upper bound on $\Delta N_{\mathrm{eff}}$ to estimate the allowed BWDM parameter space. By relating the additional DR energy density produced by annihilation to the present-day annihilation rate, we obtain an approximate constraint on the microscopic parameters, including the BWDM mass, the scalar DR mass, the coupling strength, and the condensate fraction. This analysis connects the quantum-statistical effects of the BEC to cosmological constraints on the energy transferred into relativistic dark-sector particles.

The finite-range Sommerfeld enhancement adopted here provides an analytical estimate of the low-velocity saturation regime. A more complete numerical treatment of the Yukawa two-body problem may modify the quantitative behavior of the Sommerfeld enhancement and hence the precise value of the critical condensate fraction. The present analysis is performed in a homogeneous cosmological background, which is appropriate for studying the background evolution of dark-sector energy transfer. Although annihilation can change the total BWDM particle number, we adopt approximate particle-number conservation as a zeroth-order description because the interaction is assumed to be sufficiently weak. A more complete treatment incorporating BWDM particle-number depletion and the self-consistent evolution of the thermal and condensate components would be required to investigate deviations from this approximation.

\section{Conclusion}

In this work, we have investigated the quantum statistical effects of BWDM in a microscopic dark-sector interaction model, in which energy transfer from BWDM to DR is generated by the annihilation process $\chi\chi\rightarrow\phi\phi$. The relativistic scalar particles produced in this process constitute the DR component. This framework relates the energy transfer directly to the particle-physics parameters and the phase-space distribution of BWDM.

We have considered a BWDM scenario with a mixed thermal--condensate phase-space distribution and derived the thermally averaged annihilation cross sections for the thermal--thermal, thermal--condensate, and condensate--condensate channels, including the Sommerfeld enhancement induced by the finite-range Yukawa interaction. Although the thermal--condensate channel is explicitly included, it remains subdominant in the parameter region considered here, particularly for relatively small scalar DR masses. The dominant contribution is instead governed by the competition between the thermal--thermal and condensate--condensate channels. We have identified a critical condensate fraction $r_{\rm c}$ that characterizes the transition between these two annihilation-dominated regimes. Our numerical results show that $r_{\rm c}$ remains nearly constant during the non-relativistic cosmological evolution considered in this work.

We have further investigated the constraints on the BWDM parameter space using the observational upper bound on the additional effective number of relativistic species, $\Delta N_{\mathrm{eff}}$. By relating the additional DR energy density produced by BWDM annihilation to the present-day annihilation rate, we have obtained an approximate constraint on the microscopic parameters of the model. The resulting critical BWDM mass depends on the scalar DR mass, the coupling strength $g$, and the condensate fraction $r$. In particular, the BEC component can substantially modify the relative importance of the annihilation channels. Since the condensate particles occupy the zero-momentum state, condensate--condensate annihilation probes the saturated low-velocity regime of the Sommerfeld enhancement. This enhancement can compensate for the suppression associated with the condensate number density, allowing the condensate--condensate channel to become comparable to or dominate over the thermal--thermal channel. Consequently, the presence of a BEC component can significantly modify the BWDM parameter region allowed by the observational bound on $\Delta N_{\mathrm{eff}}$.

\appendix
\section{Thermally averaged annihilation cross sections}
\label{sec:The details of thermally averaged annihilation cross sections}

In this appendix, we provide the detailed calculation of the thermally averaged annihilation cross sections for the thermal–thermal, thermal–condensate, and condensate–condensate annihilation channels. The tree-level annihilation cross section and the function $A(s)$ for the process $\chi\chi\rightarrow\phi\phi$ have been derived in Sec.\,\ref{sec:Tree-level annihilation cross section}. 
Here we focus on the phase-space averaging of the annihilation cross section over different DM components.

The thermally averaged annihilation cross section is defined as

\begin{equation}
\label{eq:thermal averaged annihilation definition appendix}
\langle\sigma\upsilon_\chi\rangle
=
\frac{
\iint
\sigma\upsilon_\chi
f_\chi(\boldsymbol p_1)
f_\chi(\boldsymbol p_2)
d^3p_1d^3p_2
}
{
\iint
f_\chi(\boldsymbol p_1)
f_\chi(\boldsymbol p_2)
d^3p_1d^3p_2
}\,.
\end{equation}

We introduce the dimensionless variables

\begin{equation}
x=\frac{M}{T}\,,
\qquad
x'=\frac{m}{T}\,,
\qquad
\Delta=\frac{\mu-M}{T}\,,
\qquad
\boldsymbol z=\frac{\boldsymbol p}{T}\,.
\end{equation}
The dimensionless BWDM distribution function becomes

\begin{equation}
\label{eq:dimensionless distribution BWDM appendix}
f_{\chi_{\rm t}}(\boldsymbol z;x,\Delta)
=
\frac{1}{(2\pi)^3}\frac{1}
{\exp(\sqrt{z^2+x^2}-x-\Delta)-1}\,,
\end{equation}
while the condensate component is described by

\begin{equation}
\label{eq:dimensionless distribution BEC appendix}
f_{\chi_{\rm c}}(\boldsymbol z;x)
=
\frac{rn_0x^3}{M^3a^3}
\delta^{(3)}(\boldsymbol z)\,.
\end{equation}

Introducing the dimensionless Mandelstam variable $s'=s/T^2$,
the function $A(s)$ appearing in the tree-level cross section becomes
\begin{equation}
\label{eq:dimensionless A appendix}
A(s')
=
4g^4
\left[
\frac{1}
{x'^4-4x'^2x^2+x^2s'}
+
\frac{
4\operatorname{arctanh}
\left(
\frac{
\sqrt{(s'-4x'^2)(s'-4x^2)}
}
{2x'^2-s'}
\right)
}
{
(2x'^2-s')
\sqrt{(s'-4x'^2)(s'-4x^2)}
}
\right]\,,
\end{equation}
where
\begin{equation}
\label{eq:dimensionless Mandelstam appendix}
s'
=
2x^2
+
2E_\chi(\boldsymbol z_1)
E_\chi(\boldsymbol z_2)
-
2z_1z_2\cos\theta\,,
\end{equation}
with $E_\chi(\boldsymbol z)=\sqrt{z^2+x^2}$.

The thermally averaged annihilation cross section can therefore be written as

\begin{equation}
\label{eq:dimensionless thermal averaged cross section appendix}
\langle\sigma\upsilon_\chi\rangle
=
\frac{x^6}{64\pi M^6}
\frac{
\iint
\frac{S}
{E_\chi(\boldsymbol z_1)E_\chi(\boldsymbol z_2)}
\sqrt{\frac14-\frac{x'^2}{s'}}
A(s')
f_\chi(\boldsymbol z_1)
f_\chi(\boldsymbol z_2)
d^3z_1d^3z_2
}
{
\iint
f_\chi(\boldsymbol z_1)
f_\chi(\boldsymbol z_2)
d^3z_1d^3z_2
}\,.
\end{equation}
The relative velocity entering the Sommerfeld factor is

\begin{equation}
\upsilon_\chi
=
\frac1x
\sqrt{
z_1^2+z_2^2-2z_1z_2\cos\theta
}\,.
\end{equation}

During the evolution from the matter-dominated epoch to the present, the BWDM particles are expected to be non-relativistic. We therefore consider the regime $x\gg1$, in which Eq.\,\eqref{eq:dimensionless distribution BWDM appendix} can be expanded as
\begin{equation}
\label{eq:nonrelativistic BWDM distribution appendix}
f_{\chi_{\rm t}}(\boldsymbol z;x,\Delta)
\simeq
\frac{1}{(2\pi)^3}\frac{1}
{\exp(z^2/2x-\Delta)-1
}\,.
\end{equation}
The accuracy of this approximation is verified by comparing the numerical results obtained from the exact and expanded distribution functions.

Introducing

\begin{equation}
y=\frac{z}{\sqrt{2x}},
\end{equation}
we obtain the annihilation cross sections for the three channels.

\begin{equation}
\label{eq:full thermal averaged annihilation cross sections}
\begin{aligned}
\langle\sigma\upsilon_\chi\rangle_{\rm tt}
=&
\frac{x^6}{128\pi M^6}
\frac{
\displaystyle
\iint dy_1dy_2
\int_{-1}^{1}d\cos\theta
\frac{S}
{E_\chi(y_1)E_\chi(y_2)}
\sqrt{\frac14-\frac{x'^2}{s'}}
A(s')
y_1^2y_2^2
f_{\chi_{\rm t}}(y_1)
f_{\chi_{\rm t}}(y_2)
}
{
\displaystyle
\iint dy_1dy_2
y_1^2y_2^2
f_{\chi_{\rm t}}(y_1)
f_{\chi_{\rm t}}(y_2)
}\,,
\\[8pt]
\langle\sigma\upsilon_\chi\rangle_{\rm ct}
=&
\frac{x^5}{64\pi M^6}
\frac{
\displaystyle
\int dy_2
\int_{-1}^{1}d\cos\theta
\frac{S|_{y_1=0}}
{E_\chi(y_2)}
\sqrt{
\frac14-\frac{x'^2}{s'|_{y_1=0}}
}
A(s'|_{y_1=0})
y_2^2
f_{\chi_{\rm t}}(y_2)
}
{
\displaystyle
\int dy_2
y_2^2
f_{\chi_{\rm t}}(y_2)
}\,,
\\[8pt]
\langle\sigma\upsilon_\chi\rangle_{\rm cc}
=&
\frac{g^6}{512\pi M^7m}
\sqrt{
\frac14
\left(
1-\frac{x'^2}{4x^2}
\right)
}\,.
\end{aligned}
\end{equation}
Here the subscripts $\rm tt$, $\rm ct$, and $\rm cc$ represent the
thermal–thermal, thermal–condensate, and condensate–condensate annihilation channels, respectively.
These expressions are used in the numerical calculations of the interaction
rate.

For analytical discussions, we further consider the limit $x\gg1$ and $x\gg x'$. In this limit, terms containing $x'$ can be neglected, and the remaining
kinematic factors are expanded in the non-relativistic regime. Therefore,

\begin{equation}
\label{eq:analytical thermal averaged cross sections appendix}
\langle\sigma\upsilon_\chi\rangle
\simeq
\begin{cases}
\dfrac{g^6}{1024\pi M^8}\xi_{\rm tt}\,, &
\mathrm{thermal–thermal}\,,
\\[8pt]
\dfrac{g^6}{512\pi M^8}\xi_{\rm ct}\,, &
\mathrm{thermal–condensate}\,,
\\[8pt]
\dfrac{g^6}{1024\pi M^7m}\,, &
\mathrm{condensate–condensate}\,,
\end{cases}
\end{equation}
where

\begin{equation}
\label{eq:xi tt appendix}
\xi_{\rm tt}
=
\frac{
\iint dy_1dy_2
\int_{-1}^{1}d\cos\theta
\dfrac{y_1^2y_2^2}
{h(\upsilon_\chi)}
f_{\chi_{\rm t}}(y_1)
f_{\chi_{\rm t}}(y_2)
}
{
\iint dy_1dy_2
y_1^2y_2^2
f_{\chi_{\rm t}}(y_1)
f_{\chi_{\rm t}}(y_2)
}\,,
\end{equation}
\begin{equation}
\label{eq:xi ct appendix}
\xi_{\rm ct}
=
\frac{
\int dy_2
\dfrac{y_2^2}
{h(\upsilon_\chi)|_{y_1=0}}
f_{\chi_{\rm t}}(y_2)
}
{
\int dy_2
y_2^2f_{\chi_{\rm t}}(y_2)}
\,,
\end{equation}
and 
\begin{equation}
h(\upsilon_\chi)=(1-e^{-2\pi\beta})\upsilon_{\chi}\,.
\end{equation}
The parameter
\begin{equation}
x=
a^2M^2
\left(
\frac{\sqrt2J^+_{3/2}(\Delta)}
{2\pi^2(1-r)n_0}
\right)^{2/3}
\end{equation}
is obtained from the approximate particle-number conservation condition,
where
$J_s^+(\Delta)
=
\Gamma(s)w_s^+(\Delta)e^\Delta
$ \cite{zhang:2024}. The function $w_s^+$ is defined as
$w_s^+(\Delta)
=
\sum_{n=0}^{\infty}
\frac{e^{n\Delta}}{(n+1)^s}$.

\section{The relation between $\Delta N_{\mathrm{eff}}$ and the present-day annihilation rate}
\label{sec:The relation between Delta N eff and the current rate of interaction}
We derive an approximate relation between the additional DR energy density produced by BWDM annihilation and the present-day annihilation rate. Starting from the continuity equation for DR, given in Eq.\,\eqref{eq:the continuity equation of DR}, the additional radiation energy density generated after matter--radiation equality can be expressed as
\begin{equation}
\Delta\rho_{\mathrm{rad}}(a)
=-\frac{1}{a^4}
\int_{a_{\mathrm{eq}}}^{a}
\frac{Q(a')a'^3}{H(a')}\,da'\,,
\end{equation}
where $Q<0$ denotes the energy transferred out of the BWDM component. Since the thermally averaged annihilation cross sections of the three channels remain approximately constant in the non-relativistic regime, and the BWDM number densities scale approximately as $a^{-3}$, we adopt $Q(a)\approx Q_0a^{-6}$. Under the matter-dominated approximation, $H(a)\approx H_0\sqrt{\Omega_m}a^{-3/2}$, the above equation becomes
\begin{equation}\label{eq:The analytical expression of the total amount of dark radiation}
\begin{aligned}
\Delta\rho_{\mathrm{rad}}(a)
&\approx-\frac{Q_0}{a^4}
\int_{a_{\mathrm{eq}}}^{a}
\frac{1}{H(a')a'^3}\,da'\\
&=-\frac{2Q_0}{H_0\sqrt{\Omega_m}\,a^4}
\left(a_{\mathrm{eq}}^{-1/2}-a^{-1/2}\right)\\
&\approx-\frac{2Q_0}{H_0\sqrt{\Omega_m a_{\mathrm{eq}}}}a^{-4},
\qquad a\gg a_{\mathrm{eq}}\,.
\end{aligned}
\end{equation}
Here, $Q_0$ denotes the present-day energy-transfer term, and $a_{\mathrm{eq}}$ is the scale factor at matter--radiation equality. The additional radiation energy density can be parameterized in terms of $\Delta N_{\mathrm{eff}}$ as \cite{Aghanim:2018}
\begin{equation}\label{eq:The parameter expression of the total amount of dark radiation}
\Delta\rho_{\mathrm{rad}}(a)
=\Delta N_{\mathrm{eff}}\times
\frac{7}{8}\left(\frac{4}{11}\right)^{4/3}
\rho_{\gamma}(a)\,,
\end{equation}
where $\rho_{\gamma}(a)=\rho_{\gamma,0}a^{-4}$ is the photon energy density. For identical BWDM particles undergoing pair annihilation, the present-day energy-transfer term is related to the present-day annihilation rate by $Q_0=-\rho_{\chi,0}\Gamma_{\chi_{\mathrm{tot},0}}$, where $\rho_{\chi,0}=\Omega_{\mathrm{BWDM}}\rho_{\mathrm{crit},0}$. Combining Eqs.\,\eqref{eq:The analytical expression of the total amount of dark radiation} and \eqref{eq:The parameter expression of the total amount of dark radiation}, we obtain
\begin{equation}
\Gamma_{\chi_{\mathrm{tot},0}}
\approx
\frac{\Omega_{\gamma}H_0\sqrt{\Omega_m a_{\mathrm{eq}}}}
{8.8\Omega_{\mathrm{BWDM}}}
\Delta N_{\mathrm{eff}}\,,
\end{equation}
where $\Omega_{\gamma}\approx5.35\times10^{-5}$ denotes the density parameters of photons.

\acknowledgments

Zhijian Zhang thanks Weikang Lin and Zhengxiang Li for helpful discussions
and Kevin J. Ludwick for clarification regarding the temperature parameter in
the dark matter distribution function. This work was supported by the National Key Research and Development Program of China Grant Nos. 2023YFC2206702, and 2021YFC2203001; National Natural Science Foundation of China under Grants Nos. 11920101003, 12021003, 11633001, 12322301, and 12275021; the Strategic Priority Research Program of the Chinese Academy of Sciences, Grant Nos. XDB2300000 and the Interdiscipline Research Funds of Beijing Normal University.




\bibliography{biblio}
\bibliographystyle{JHEP}

\end{document}